\documentclass[twoside,fleqn]{article}

\usepackage{epsf}
\usepackage{espcrc2}
\usepackage{amstex}
\usepackage{amssymb}

\title{Fixed-point pure gauge action using $ b= \sqrt{3} $ RGT\thanks{
        This work was supported by the DoE Grand Challenges award 
	at the ACL at Los Alamos.}
        }
\author{Tanmoy Bhattacharya\address{MS B-285, 
		Los Alamos National Lab, Los Alamos, New Mexico 87545, USA}, 
	Rajan Gupta${}^{\rm a}$, 
	Weonjong Lee${}^{\rm a}$, 
       }

\begin{document}

\begin{abstract}
We present a status report on the construction of the classical
perfect action using the $ b=\sqrt{3} $ renormalization group
transformation (RGT)~\cite{gupta-1}.  
We investigate finite volume corrections and map the locality of the
fixed-point action by tuning the RGT parameter, $\kappa$. 
We compare results with the previous calculation for $b=2$ RGT~\cite{DHHN-1}. 
\end{abstract}

\maketitle

\section{$b=\sqrt 3$ RGT}

The $b=\sqrt 3$ RGT~\cite{gupta-1} has a number of advantages compared
to $b=2$~\cite{DHHN-1,wiese-1}: 
it 1) has a smaller step size, 2) incorporates
more gluonic degrees of freedom, 3) requires less tuning parameters,
4) has no overlap of either gauge or fermion fields between
neighboring block cells, and 5) preserves a higher rotational symmetry
for matter fields. This work complements the previous estimate of the
renormalized action generated at $1/a \sim 2$ GeV using
MCRG~\cite{gupta-2}.

The orthogonal basis vectors defining the block lattice 
are the body-diagonals of four positively oriented cubes
$e^{(1)} =  T_1 \cdot e^{(0)} $, where $T_1$ is chosen to be 
\begin{eqnarray}
& & \hspace*{-7mm}
\left( \begin{array}{c} 
	 e^{(1)}_x \\ e^{(1)}_y \\ e^{(1)}_z \\ e^{(1)}_t
	\end{array} \right) =
	\left( \begin{array}{cccc}
		1  &  1 & 1  & 0 \\
		-1 &  1 & 0  & 1 \\
		1  &  0 & -1 & 1 \\
		0  & -1 & 1  & 1 
		\end{array} \right)
	\left( \begin{array}{c} 
	e^{(0)}_1 \\ e^{(0)}_2 \\ e^{(0)}_3 \\ e^{(0)}_4
	\end{array} \right)
\label{basis-1}
\end{eqnarray}
The block lattice has twisted boundary conditions (TBC), 
(the block lattice obtained from $ (3L)^4 $ periodic lattice
is a $ (3L)^2 \times L^2 $ lattice with TBC)
\begin{eqnarray}
& & f ( x+3L, y, z, t)      = f (x,y,z,t)  \nonumber \\
& & f ( x, y+3L, z, t)      = f ( x, y, z, t) \nonumber \\
& & f ( x, y, z+L, t)       =  f (x-L,y+L,z,t) \nonumber \\
& & f ( x, y, z, t+L)       =  f ( x+L, y+L, z, t)
\label{tbc-st}
\end{eqnarray}
We choose the second transformation to be
\begin{eqnarray}
e^{(2)} &=& T_2 \cdot e^{(1)} = 3 e^{(0)} \\
T_2 & \equiv & (T_1)^T \mbox{ such that }  T_2 \cdot T_1 = 3
\end{eqnarray}
and we iterate these two steps.

The momentum space on periodic lattices is 
\begin{eqnarray}
p^{(0)}_\mu = \frac{2\pi n} {3L} \mbox{ with } n \in \{1,2, \cdots, 3L\}
\\
p^{(2)}_\mu = \frac{2\pi m} {L} \mbox{ with } m \in \{1,2, \cdots, L\}
\end{eqnarray}
whereas, for the twisted lattice the Brillouin zone is defined as
\begin{eqnarray}
p^{(1)}_x &=& \frac{2\pi n}{3L}
\, ; \quad \qquad p^{(1)}_y = \frac{2\pi m}{3L}
\nonumber \\
p^{(1)}_z &=& p^{(1)}_y - p^{(1)}_x + \frac{2\pi l}{L}
\nonumber \\
p^{(1)}_t &=& p^{(1)}_y + p^{(1)}_x + \frac{2\pi k}{L}
\label{mom-tbc}
\end{eqnarray}
where $ n,m \in \{1,\cdots,3L\} $ and $l,k \in \{1, \cdots, L \} $.

\begin{figure}[!t]
\vskip -2mm
\epsfxsize=\hsize
\epsfbox{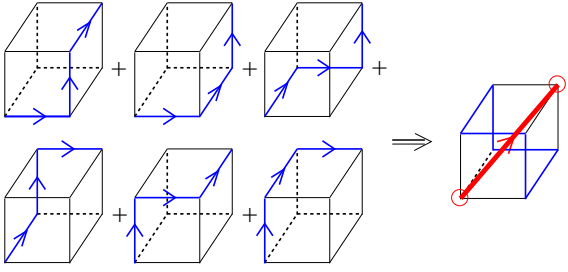}
\vskip -8mm
\caption{$b=\sqrt{3}$ blocking transformation.}
\vskip -7mm
\label{fig:block}
\end{figure}

The momentum on the coarse lattice is connected with
that on the fine lattice as follows:
\begin{eqnarray}
p^{(i)} &=& \frac{p^{(i+1)}} {\sqrt{3}} + h^{(i+1)}, 
\mbox{ with }
p^{(j)} \equiv \sum_{\mu} p^{(j)}_\mu \frac{ e^{(j)}_\mu }{a^{(j)}}
\nonumber \\
h^{(j)} &=& \frac{ 2\pi } {3}
        \bigg( \frac{e^{(j)}_z}{a^{(j-1)}} h_z +
                \frac{e^{(j)}_t}{a^{(j-1)}} h_t \bigg)
\label{h-mom}
\end{eqnarray}
where $ a^{(i)}$ is the lattice spacing at 
$i$-th iteration, and $ h_z, h_t \in \{0,1,2\} $.
The block link $Q$ is constructed 
from the average of the 6 independent 3-link 
paths connecting the body-diagonal as shown in Fig.~\ref{fig:block}
\begin{eqnarray}
& & \hspace*{-5mm}
	Q^{(1)}_x(n^{(1)}) = \frac{1}{6} \ [
        U_1(n^{(0)}) U_2(n^{(0)} + \hat e^{(0)}_1)
\nonumber \\        
& & \hspace*{-5mm}
	U_3(n^{(0)} + \hat e^{(0)}_1 + \hat e^{(0)}_2) +
	\mbox{ (123) permutation } ] 
\end{eqnarray}
The fixed-point (FP) equation 
as $ \beta \rightarrow \infty $~\cite{DHHN-1} is:
\begin{eqnarray}
& & \hspace*{-7mm} S^{FP}(V) =\min_U ( S^{FP}(U) + T(U,V) )
\label{eq:fp} \\
& & \hspace*{-7mm} T(U,V) \equiv   -
        \frac{\kappa}{N} \sum_{n^{(1)}}
        [ {\rm Re} {\rm Tr} ( V_m Q^{(1)}_m {}^\dagger ) - F(U) ]
\\
& & \hspace*{-7mm} 
F(U) \equiv \max_{W} \{ {\rm Re} {\rm Tr} ( W_m Q^{(1)}_m {}^\dagger ) \}
\nonumber
\end{eqnarray}
%
%
%
The parameter $\kappa$ is tuned to optimize the locality of the
FP action.  Expanding the FP action to quadratic order in the
gluon fields gives 
\begin{eqnarray}
& & \hspace*{-10mm} \frac{1}{V^{(1)}}  \sum_{k^{(1)}}
\rho^{(1)}_{mn}( k^{(1)} ) {\rm Tr} [ B_m^* (k^{(1)}) B_n (k^{(1)}) ]
=
\nonumber \\
& & \min_{A} \bigg(
\frac{1}{V^{(0)}} \sum_{k^{(0)}} 
	\rho^{(0)}_{\mu\nu} {\rm Tr} [ A_\mu^* (k^{(0)}) A_\nu (k^{(0)}) ] +
\nonumber \\
& & \frac{\kappa} {V^{(1)}} \sum_{ k^{(1)} }
        {\rm Tr} \mid \Gamma_m (k^{(1)}) - B_m (k^{(1)}) ) \mid^2 
\bigg)
\label{eq:fp-2}
\end{eqnarray}
$ B_m $ and $ A_\mu $ are gluon fields on the coarse and
fine lattice respectively, $ k^{(0)} = (k^{(1)} / \sqrt{3}) + h^{(1)} $,
$ V^{(0)} = 9 V^{(1)} $. We use 
Greek indices for periodic lattice and 
Roman indices for the twisted lattice.  Lastly, 
\begin{eqnarray}
\Gamma_m (k^{(1)}) &\equiv& \frac{1}{9}\sum_{ h^{(1)} }
\omega^{(1)}_{m\nu} (k^{(0)}) A_\nu (k^{(0)})
\\
\omega^{(1)}_{x1} (k) &=& \frac{1}{4}
        \bigg( \frac{ 2 \hat{k}_2  } { \hat {k}_2 } \bigg)
        \bigg( \frac{ 2 \hat{k}_3  } { \hat {k}_3 } \bigg) +
        \frac{1}{12} \hat {k}_2 \hat {k}_3
\end{eqnarray}
and similar expressions for the other components of $\omega^{(1)}$. 
The matrices $ \omega^{(1)} $ and vectors $ h^{(1)} $ 
contain all the details of the RGT.
Solving Eq. (\ref{eq:fp-2}) leads to the recursion relation 
for $D^{(i)} \equiv ( \rho^{(i)} )^{-1}$:
\begin{eqnarray}
& & \hspace*{-7mm} D^{(1)}_{ml} ( k^{(1)} ) = 
\nonumber \\
& & \hspace*{-7mm} 
\frac{1}{9} \sum_{ h^{(1)} }
        \omega^{(1)}_{m \mu} (k^{(0)})
        D^{(0)}_{\mu\beta} (k^{(0)}) 
	\omega^{(1)}_{\beta l}{}^\dagger (k^{(0)})
        + \delta_{ml} \frac{1}{\kappa}
\label{eq:fp-rr}
\end{eqnarray}
Similarly, the recursion relation
for $ T_2 $ is obtained by changing 
$ (0) \rightarrow (1) $, $ (1) \rightarrow (2) $, and 
$ \mu ({\rm Greek}) \leftrightarrow m ({\rm Roman})$.
\begin{eqnarray}
\omega^{(2)}_{1 x}(k) &=&  \frac{1}{4}
        \bigg( \frac{ 2 \hat{ k}_y  } { \hat {k}_y } \bigg)^*
        \bigg( \frac{ 2 \hat{ k}_z  } { \hat {k}_z } \bigg) +
        \frac{1}{12} \hat {k}_y^* \hat {k}_z
\end{eqnarray}

The specific choice of $ T_1 $ given in Eq. (\ref{basis-1}) is not
unique.  There are eight equivalent independent choices. So for each
$T_2 \otimes T_1$, we average over the eight to regain hypercubic
invariance.

\section{Numerical study of FP action}

To find the FP we start with $ D^{(0)}$ equal to the free propagator
in Feynman gauge and numerically iterate the recursion relations
Eqs.~(\ref{eq:fp-rr}) until $ D^{(2n)} $ satisfies the
fixed-point requirement: for a given precision criterion $\epsilon$
and $ \forall k $ in the $(2n)$-th Brillion zone,
\begin{eqnarray}
\mid D^{(2n)}_{\mu\nu} (k) - D^{(2n-2)}_{\mu\nu} (k) \mid < \epsilon.
\end{eqnarray}
%

%
%
\begin{figure}[!t]
\vspace*{-2mm}
\epsfxsize=\hsize\epsfbox{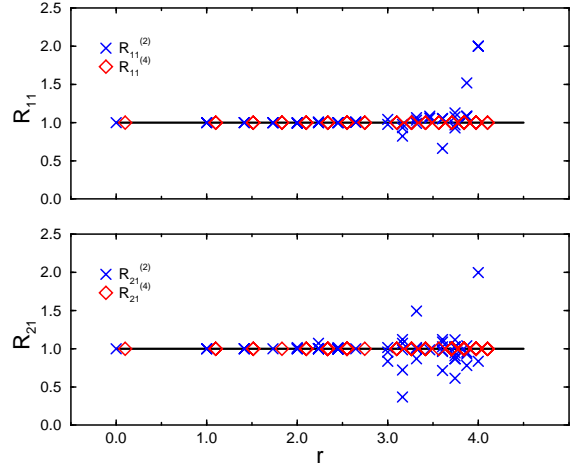}
\vskip -12mm
\caption{Finite volume effect.}
\vskip -7mm
\label{fig:vol}
\end{figure}
We now discuss results. 
First, in Fig. \ref{fig:vol}, we show the ratios:
\begin{eqnarray*}
& & R_{\mu\nu}^{(2)} = \frac{ \rho_{\mu\nu} (L: 24 \rightarrow 8 ) }
		{ \rho_{\mu\nu} (L: 216 \rightarrow 72 ) } 
\\ 
& & R_{\mu\nu}^{(4)} = \frac{ \rho_{\mu\nu} (L: 72 \rightarrow 24 ) }
		{ \rho_{\mu\nu} (L: 216 \rightarrow 72 ) } 
\end{eqnarray*}
to show finite volume corrections. On the $8^4$ lattice, we observe
$\ll 1 \%$ deviation for $ r \equiv \vert x \vert \le 2 $, 
which grow significantly for $ r \gtrsim 3 $. 
Since no correction is observed on the $ 24^4 $
lattice, we assume that $\rho (L: 216 \rightarrow 72 )$ gives infinite
volume results in the range $ r \lesssim 4 $, and
$24^4$ coarse lattice is sufficient for the calculation.
%

%
%
\begin{figure}[!t]
\vskip -2mm
\epsfxsize=\hsize\epsfbox{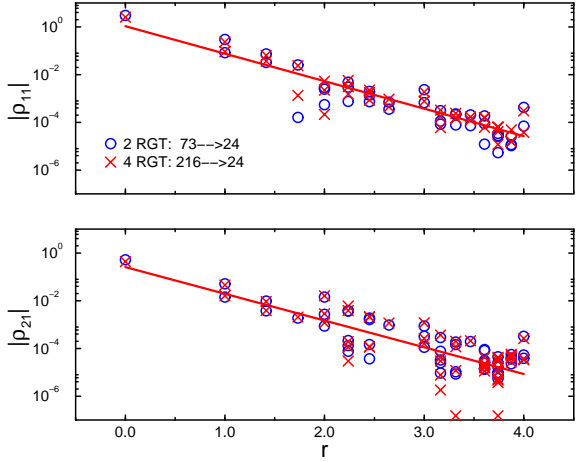}
\vskip -12mm
\caption{Minimal $\chi^2$ fitting  at $\kappa=12.0$.}
\vskip -7mm
\label{fig:fit}
\end{figure}
\begin{figure}[!t]
\vskip -2mm
\epsfxsize=\hsize\epsfbox{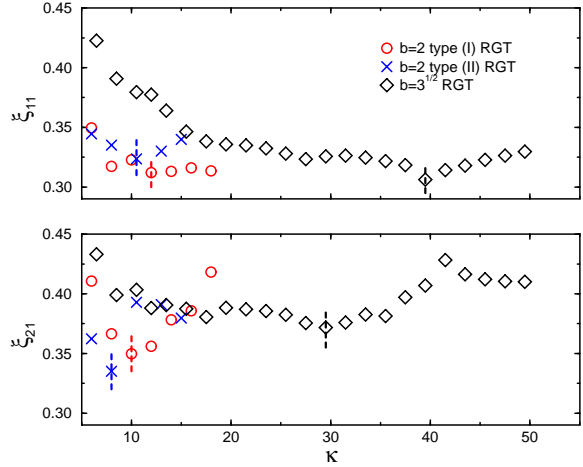}
\vskip -12mm
\caption{$\xi_{\mu\nu}$ with repect to $ \kappa $.}
\vskip -7mm
\label{fig:locality}
\end{figure}

Second, we optimize $\kappa$ for maximal locality in 
both in $\rho_{11}$ and $\rho_{21}$ couplings. 
We define the locality parameter $ \xi $ as
\begin{eqnarray}
\mid \rho_{\mu\nu} (r) \mid = Z \exp\left( - \frac{ r }{ \xi_{\mu\nu} } \right)
\label{eq:loc-fit}
\end{eqnarray}
Fig. \ref{fig:fit} shows a least square fit which determines $\xi$. 
Since the fit is very sensitive to $ r \geq 2 $, we only 
use the $ 24^4 $ lattice results obtained by 4 RGT iterations 
to avoid finite volume effects.
The maximal locality is found at $ \kappa = 39.5 $ for 
$\xi_{11}$ and $ \kappa = 29.5 $ for 
$\xi_{21}$ as shown in Fig. \ref{fig:locality}. 
%
%
\begin{table}[!h]
\vskip -7mm
\begin{center}
\begin{tabular}{ c@{\hspace{1mm}} c@{\hspace{1mm}} c@{\hspace{1mm}}
	|@{\hspace{1mm}} 
	c@{\hspace{1mm}} c@{\hspace{1mm}} c@{\hspace{1mm}} }
\hline
\multicolumn{3}{ c@{\hspace{1mm}}|@{\hspace{1mm}} }{$\rho_{00}(r)$} 
					& \multicolumn{3}{c}{$\rho_{10}(r)$}	\\
\hline
r	& $b=\sqrt{3}$ 	& $ b=2 $ (I)	& r	&  $b=\sqrt{3}$ & $ b=2 $ (I) 	\\
\hline					
0001	& $ -0.4110 $	& $-0.6718$	& 0000	& $ -0.6977 $	& $-0.8007$	\\ 
1000    & $ -0.2909 $	& $-0.0595$	& 0001	& $ -0.0686 $	& $-0.0310$	\\
0011    & $ -0.1027 $	& $-0.0428$	& 0100  & $ +0.0524 $	& $+0.0113$	\\
1001    & $ +0.0645 $	& $+0.0161$	& 0101  & $ -0.0019 $	& $-0.0021$	\\
0111    & $ -0.0336 $ 	& $-0.0272$	& 1011  & $ +0.0126 $	& $+0.0088$	\\
1011    & $ +0.0053 $  	& $-0.0027$	& 2100  & $ +0.0109 $	& $-0.0039$	\\
1111	& $ -0.0020 $	& $-0.0037$	& 0111  & $ -0.0020 $	& $-0.0010$	\\
0002    & $ +0.0145 $  	& $+0.0242$	& 0002  & $ +0.0766 $	& $-0.0017$	\\
				
\hline
\end{tabular}
\end{center}
\caption{ $\rho_{\mu\nu}(r)$ after 6 iterations of $b=\sqrt{3}$ RGT.}
\label{tab:rho}
\vspace{-7mm}
\end{table}
Our optimal choice is taken to be $\kappa = 27.5 $ and we present 
the corresponding $\rho_{\mu\nu}$ in Tab. \ref{tab:rho}. For comparison 
we also give the results for $b=2$ RGT first obtained in~\cite{DHHN-1}. 
We are not able to comment on which RGT is ``better'' due to the issue 
of redundant operators; numerical tests of scaling need to be done. 

Finally, the results for the couplings given in Tab.~\ref{tab:rho} are
expressed in terms of coefficients of Wilson loops in
Tab.~\ref{tab:WL}.  Simulations to check the efficacy of this action
need to be done.
%
%
%
%
\begin{table}[h]
\vskip -7mm
\begin{center}
\begin{tabular}{ l | c }
\hline
Wilson Loop 		&	coefficients 	\\ 
\hline					
(x,y,-x,-y)		&	$ +0.6285 $ 	\\ 
\hline					
(x,y,y,-x,-y,-y)	&	$ -0.0830 $ 	\\ 
\hline					
(x,y,z,-x,-y,-z)	&	$ +0.0419 $ 	\\ 
\hline					
(x,y,x,z,-x,-y,-x,-z)	&	$ -0.0041 $ 	\\ 
\hline					
(x,y,y,-x,z,-y,-y,-z)	&	$ +0.0060 $ 	\\ 
\hline					
(x,y,y,-x,-x,-y,-y,x)	&	$ +0.0105 $  	\\ 
\hline					
(x,y,z,t,-x,-y,-z,-t)	& 	$ +0.0062 $	\\ 
\hline
\end{tabular}
\end{center}
\caption{Coefficients of Wilson loops in the FP action.}
\label{tab:WL}
\vspace{-7mm}
\end{table}
%



%

\end{document}